\documentclass[10pt,
               aps,
               prl,
               twocolumn,
               showpacs,
               groupedaddress]{revtex4-2}
\usepackage[utf8]{inputenc}
\usepackage{graphicx}
\usepackage[caption=false]{subfig}
\usepackage{amsmath, amsfonts, amssymb}
\usepackage{float}
\usepackage[T1]{fontenc}  
\usepackage[linktocpage=true,
  colorlinks=true, 
  pdfborder={0 0 0},
  linkcolor=blue,
  citecolor=blue,
  filecolor=yellow,
  urlcolor=blue,
  bookmarks,
  pdfauthor={},
]{hyperref}
\usepackage{tikz}

\makeatletter
\newcommand*{\rom}[1]{\expandafter\romannumeral #1}
\makeatother
\begin{document}

\title{TUR Violation, Minimal Power Fluctuation, and Enhanced Efficiency}

\author{F. Adersh}

\author{M. Sahoo}
\email{jolly.iopb@gmail.com}
\affiliation{Department of Physics, University of Kerala, Kariavattom, Thiruvananthapuram-$695581$, India}

\date{\today}

\begin{abstract}
We investigate the performance characteristics of an active magneto gyrator that consists of an inertial charged active particle confined in a two-dimensional asymmetric parabolic potential, coupled to two thermal baths maintained at two different temperatures and in the presence of a magnetic field applied perpendicular to the plane of motion. In contrast to the passive Brownian gyrator, where the power-efficiency trade-off imposes strict operational limits, the active magneto gyrator exhibits striking violations of this rule, allowing regimes in which both output power and efficiency increase simultaneously with load strength. Moreover, while the Thermodynamic Uncertainty Relation (TUR) predicts that highly efficient engines necessarily produce strongly fluctuating power, rendering the Brownian gyrator unusable as the efficiency approaches the Carnot bound, the active magneto gyrator circumvents this constraint. Specifically, it achieves high efficiency with minimal power fluctuations, thus defying the TUR.  Remarkably, the efficiency can even surpass the Carnot limit and approach 100\%, without compromising stability. This condition cannot be achieved in the absence of either activity or magnetic field. These results highlight the active magneto gyrator as a promising setup for designing stochastic heat engines that can
transcend Carnot efficiency.
\end{abstract}

\maketitle
\section{INTRODUCTION}\label{sec:intro}

Brownian heat engines are microscopic thermodynamic systems that convert thermal energy into mechanical work~\cite{ciliberto2017experiments,martinez2015adiabatic,park2016efficiency,Chiang2017electrical,martines2017colloidal,sanjib2017stochastic,anupam2019nonequilibrium,Lee2020Brownian,Chang2021autonomous,abdoli2022tunable,Lin2022Stochastic,han2025TUR}. Unlike macroscopic engines, thermal noise plays a dominant role in the dynamics of Brownian engines, leading to fluctuations in the thermodynamic currents~\cite{Holubec2022fluctuations}. As a result, the classical power-efficiency trade-off rule which establishes the fundamental impossibility of simultaneously achieving high efficiency and finite power output in heat engines, has to be extended to incorporate power fluctuations, thereby linking power, efficiency, and fluctuations within a unified framework~\cite{Pietzonka2018universal}. This complex interplay can be described with the help of the thermodynamic uncertainty relation (TUR). In the recent years, this TUR is emerged as a fundamental framework linking the precision of fluctuating thermodynamic currents such as output power and heat to the rate of entropy production in nonequilibrium systems~\cite{Seifert2012stochastic,Barato2015Thermodynamic,gingrich2016dissipation,gingrich2017fundamental,horowitz2017proof,Horowitz2020Thermodynamic,Li2019quantifying}. It implies that achieving highly precise power output necessarily incurs a finite dissipation cost, thereby imposing universal constraints on the simultaneous optimization of power, efficiency, and fluctuations in stochastic heat engines. Consequently, steady-state engines obeying the TUR cannot approach Carnot efficiency at finite power~\cite{Pietzonka2018universal,Lin2022Stochastic}, which imposes stringent limits on engine performance and reliability.

Beyond the realm of passive thermal systems, the field of active matter comprising self-propelled particles that consume environmental energy to generate persistent motion has opened new avenues of research especially in nonequilibrium statistical mechanics and bio-physics~\cite{ramaswamy2017active,Maggi2014generalized,magistris2015intro,bechinger2016active,deblais2018boundaries,dauchot2019dynamics, Gompper2020roadmap,pietzonka2021oddity}. Incorporating activity into microscopic engines can fundamentally alter the stochastic energetics and fluctuation properties, thereby enhancing performance compared to passive systems~\cite{kumari2024Stochastic, saha2019stochastic,Datta2022Secondlaw,datta2022second}. Motivated by these developments, active extensions of microscopic Brownian engines have been explored, in which active  systems operate as heat engines with distinct performance characteristics, including modified trade-off relations between power and  efficiency~\cite{Lee2020Brownian}. Additionally, active heat engines are not constrained by a universal efficiency bound and can achieve efficiencies arbitrarily close to unity~\cite{pietzonka2019Autonomous,Adersh2025heatpump}. The Brownian gyrator is a minimal stochastic model of an autonomous heat engine, comprising a Brownian particle in an anisotropic potential coupled to two heat baths maintained at different temperatures~\cite{filliger2007brownian, dotsenko2013two, mancois2018two, Chiang2017electrical, bae2021inertial, Argun2017Experimental, Chang2021autonomous, baldassarri2020engineered, cerasoli2021time, fogedby2017minimal, cerasoli2022spectral}. The resulting temperature gradient and potential anisotropy generate steady-state rotational currents that manifest themselves as a systematic torque, which can be harnessed for work extraction~\cite{park2016efficiency, abdoli2022tunable, Chiang2017electrical, Chang2021autonomous,Lin2022Stochastic}. This mechanism has also been experimentally demonstrated~\cite{Argun2017Experimental, Chiang2017electrical}. 

Moreover, such a system in the presence of an external magnetic field is found to work as a tunable heat pump~\cite{abdoli2022tunable} and shows nontrivial steady state properties due to Lorentz force induced spatial correlations, while coupled to multiple heat baths~\cite{abdoli2020correlations}. Interestingly, such a two temperature setup with an active Ornstein-Uhlenbeck particle~\cite{szamel2014self,Nguyen2021ActiveOM,caprini2021inertial,muhsin2021orbital,muhsin2022inertial,muhsin2023memory,muhsin2023ratchet,arsha2023steady,adersh2024inertial,adersh2025transition,sanju2026Generalized,deion2026chiral,uma2026active} while confined by an asymmetric harmonic trap and in the presence of magnetic field gives rise to emerging behavior like memory induced trapped diamagnetism~\cite{Muhsin2025Active} in a viscoelastic environment and is observed to achieve remarkably high efficiency in the engine mode~\cite{Adersh2025heatpump}. It is further interesting to investigate the thermodynamic performance characteristics of such a system by analyzing the thermodynamic quantities and the associated fluctuation in the output power in order to assess their practical viability as microscopic heat engines. Detailed analysis of fluctuations provides valuable insight into the stability and reliability of the engine, allowing the validity of the TUR~\cite{Pietzonka2018universal}. Furthermore, evaluating the entropy production rate is essential for characterizing the irreversibility of the engine and determining the thermodynamic cost of its operation~\cite{Lee2020Brownian,Lin2022Stochastic}.

Motivated by this, in this work, we have investigated the thermodynamic performance characteristics and associated power fluctuations of an active magneto gyrator composed of an inertial active Ornstein-Uhlenbeck particle confined in a two-dimensional asymmetric parabolic potential, coupled to two thermal baths maintained at different temperatures and in the presence of magnetic field. In the presence of a temperature gradient and potential asymmetry, the active particle undergoes gyration across the potential minimum and exhibits precession due to the presence of magnetic field. The effective gyration due to the combined effects of both gyration and precession mechanisms results in a nonzero average torque on the confining potential, as expected. Subjection of an external load opposite to the direction of the induced torque enables extraction of work from the system, allowing it to operate as a heat engine. Within this framework, we have exactly computed the thermodynamic performance characteristics and also the associated output power fluctuations. Interestingly, we observed that the efficiency of the system can approach the Carnot limit and may even reach unity. Moreover, violation of the conventional power-efficiency trade-off~\cite{Pietzonka2018universal,Lee2020Brownian} is observed, allowing regimes in which the system achieves both maximum power and maximum efficiency simultaneously at the same load strength. We also find that achieving efficiencies equal to or exceeding the Carnot efficiency is necessarily accompanied by a violation of the TUR. For a typical heat engine that satisfies the TUR, Carnot efficiency represents the maximum attainable efficiency and is achieved only when entropy production rate and output power becomes zero, while the power fluctuations diverge~\cite{Lin2022Stochastic,Pietzonka2018universal}. However, in our model, interestingly, we have identified a novel parameter regime where the system achieves both high power and high efficiencies, including Carnot and unit efficiency, while exhibiting minimal power fluctuations despite vanishing or negative entropy production rates, thereby violating the TUR. This doesn't occur in the absence of either magnetic field or activity. Hence, active magneto gyrator can work as an ideal heat engine that can surpass Carnot efficiency with minimal power fluctuation while maintaining stability unlike a Brownian magneto gyrator or an active gyrator.

In the next section, we introduce the model under consideration, followed by a discussion of the main results, and finally conclude. 

\section{MODEL}\label{sec:model}
We consider the stochastic motion of an inertial charged active particle of mass $m$ confined within a two-dimensional (2D) asymmetric harmonic potential defined by $V(x,y)=\frac{1}{2}k \left(x^2+y^2\right)+\alpha x y$, where $k$ is the trap stiffness and $\alpha$ characterizes the degree of asymmetry of the potential. The particle carries a charge $q$ and is subjected to a Lorentz force induced by a uniform magnetic field ${\bf B}=B \hat{k}$ applied perpendicular to the plane of motion. We describe the system dynamics using the Langevin model \cite{caprini2021inertial,Nguyen2021ActiveOM,Muhsin2025Active,Adersh2025heatpump} given by
\begin{equation} 
m\ddot{{\bf r}}(t) =-\gamma \dot{{\bf r}}(t)+|q|\left({{\bf{\dot{r}}}(t)\times{\bf B}}\right) - \nabla V +{\bf F}^{nc}+{{\boldsymbol{\xi}}}(t)+ {{\boldsymbol{\zeta}}}(t),
\label{eq:maineqmotion-vector}
\end{equation}
where ${\bf r}(t)=x(t)\hat{i}+y(t)\hat{j}$, denotes the position vector of the particle. The force ${\bf F}^{nc} = \varepsilon( {\bf r}(t) \times \hat{k})$ acts perpendicular to the position vector, with the parameter $\varepsilon$ controlling its strength. This force is nonconservative since $\nabla\times {\bf F}^{nc}\neq0$ and for $\varepsilon > 0$ generates a torque directed along the negative $z$-direction. $\boldsymbol{\zeta}(t)$ represents the thermal noise force with zero mean and correlation given by $\langle\zeta_i(t)\zeta_j(t')\rangle=2\delta_{ij} \gamma T_i\delta (t-t')$, with $(i,j) \in \{x, y\}$, corresponding to $x$ and $y$ degrees of freedom in the $xy$-plane. $T_c = T_x$ and $T_h = T_y$ denote the temperatures of the cold and hot baths, respectively.
The term $\boldsymbol{\xi}(t)$ represents the active force and it follows the Ornstein-Uhlenbeck (OU) process
\begin{equation}
t_c \dot{\boldsymbol{\xi}} = -\boldsymbol{\xi} + \xi_0\sqrt{2 t_c} \boldsymbol{\mu}(t).
\label{eq:noise}
\end{equation}

Here, $\boldsymbol{\xi}(t)$ is the Gaussian process with zero mean and characterized by an exponentially decaying time correlation
\begin{equation}
    \langle \xi_i(t) \rangle = 0; \qquad \langle \xi_i(t) \xi_j(t') \rangle = \delta_{ij} \xi_0^2 \exp{\left( \frac{|t - t'|}{t_c} \right)}.
    \label{eq:noise_stat}
\end{equation}
The time constant $t_c$ denotes the persistence duration of the particle's self-propelled motion, thus serving as the activity timescale. The quantity $\xi_0$ denotes the strength of the active force, while $\boldsymbol{\mu}(t)$ represents a delta-correlated Gaussian white noise vector. 

The TUR establishes a fundamental constraint that binds the power fluctuations to its thermodynamic cost. The TUR in the steady state far from equilibrium is given by the relation~\cite{Pietzonka2018universal,Lin2022Stochastic}
\begin{equation}
     \tilde{{\bf J}} \sigma\ge k_B,
     \label{eq:TUR_main}
 \end{equation}
where $\sigma$ denotes the entropy production rate, $
{\bf \tilde{J}}$ represents the relative fluctuation in the output power and $k_B$ is the Boltzmann constant, which is set to unity ($k_B$=1) throughout the paper. Entropy production plays a central role in assessing both the validity and the tightness of the TUR. Within this framework, $\sigma$ quantifies the thermodynamic cost required to sustain nonequilibrium currents. Consequently, it establishes a direct relation between heat dissipation and the achievable precision of the output power. The entropy production rate is given by~\cite{Lee2020Brownian,Lin2022Stochastic}
\begin{equation}
     \sigma=\frac{\Sigma}{t} =-\frac{Q_h}{T_h}-\frac{Q_c}{T_c},
     \label{eq:EPR_main}
\end{equation}
where $\Sigma$ represents the entropy production. $Q_h$ and $Q_c$ denote the steady-state heat dissipation rates from the hot and cold baths, respectively, which can be calculated as
\begin{equation}
 Q_h=\Bigl \langle \frac{\textit{đ}\mathcal{Q}_y}{dt}\Bigr \rangle,\quad\text{and} \quad Q_c=\Bigl \langle \frac{\textit{đ}\mathcal{Q}_x}{dt}\Bigr \rangle.
 \label{eq:Heat_avg_main}
\end{equation}

Within the framework of stochastic thermodynamics, the heat exchanged between the particle and the cold and hot baths associated with the $x$ and $y$ degrees of freedom, denoted by the parameters \textit{đ}$\mathcal{Q}_x$ and \textit{đ}$ \mathcal{Q}_y$, are given by~\cite{sekimoto1997langevin, sekimoto2010stochastic}

\begin{equation}
    \textit{đ}{\mathcal{Q}}_i=v_i\circ\bigr[-\gamma  v_i(t)+{ {\xi}_i}(t)+{ {\zeta}_i}(t)\bigr],
    \label{eq:dQh}
\end{equation} 
where $v_i$ is the velocity component along the $i$-axis, with $i=(x,y)$. The symbol ``$\circ$'' denotes multiplication in the Stratonovich sense. 

For a heat engine operating in the steady state, the entropy production rate and the extracted power are directly linked to the efficiency. Consequently, one can obtain $\sigma$ in the following form starting from Eq.~\eqref{eq:EPR_main}
\begin{equation}
\sigma \;=\; 
\frac{P}{T_c}\left(\frac{\eta_c}{\eta} - 1\right),
\label{eq:EPR-withEff}
\end{equation}
 where $\eta_c$ is the Carnot efficiency and $P$ is the average power that can be obtained by applying a constant load $\mathbf{F}^{nc}$ in the direction opposite to the induced torque of the gyrating particle~\cite{abdoli2022tunable,Adersh2025heatpump}. The average power delivered by the system in the steady state is given by 
\begin{equation}
     P=\langle {\bf F}^{nc}\cdot {\bf \dot{r}}\rangle=\varepsilon\langle x \dot{y}-y \dot{x} \rangle.
     \label{eq:power_main}
\end{equation}

Similiarly, the power extracted at a finite time $t$ is given by
\begin{equation}P(t) = \frac{\varepsilon}{t}  \int_{0}^{t} \left[ x(t') \circ \dot{y}(t')-y(t')\circ \dot{x}(t') \right] \, dt'.
\end{equation}

Now, relating the average power $P$ with the corresponding heat input from the hot bath $Q_h$,  the efficiency $\eta$ can be defined as
 \begin{equation}
    \eta=\dfrac{P}{Q_h}.
    \label{eq:Eff_main}
\end{equation}

The power fluctuations describe the stochastic deviation of power and determine the precision of transport processes. The relative fluctuations of the output power in the steady state is given by~\cite{Pietzonka2018universal,Lin2022Stochastic}
\begin{equation}
    \tilde{{\bf J}} = \frac{2 D_P}{P ^2},
    \label{eq:TUR_Jtilde}
\end{equation}
with $D_P$ as the power fluctuations and it is given by
\begin{equation}
D_P= \displaystyle\lim_{t \to \infty} 
\langle \left( P(t) -  P  \right)^2\rangle t.
\label{eq:dw_main}
\end{equation}

We numerically simulate the dynamics governed by Eq.~\eqref{eq:maineqmotion-vector} together with the evolution of the active force described by Eq.~\eqref{eq:noise} using the Euler--Maruyama method~\cite{Kloeden1992numerical}. The simulation is performed with a time step of $10^{-3}$ and $10^{6}$ time steps are used to ensure that the system reaches steady state. The ensemble averages of the physical quantities are subsequently computed over $10^{7}$ independent realizations.
 
 \section{RESULTS AND DISCUSSION}\label{sec:result}
In order for solving the dynamics, we recasted the dynamics of the particle governed by Eqs.~\eqref{eq:maineqmotion-vector} and \eqref{eq:noise} into an equivalent set of three Markovian equations as follows:
 \begin{equation}
 \begin{aligned}
    &{\bf \dot{r}} = {\bf v}, \\
    &{\bf \dot{v}} = -{\Gamma}{ \bf v} -\frac{\nabla V}{m} + \left(\omega_c{\bf v} + \frac{\varepsilon}{m} {\bf r}\right) \times \hat{k} + \frac{{\boldsymbol \xi}}{m} + \frac{{{\boldsymbol{\zeta}}}}{m} ,\\ 
    &\text{and}\qquad  
      {\dot{\boldsymbol \xi}} = -\frac{{\boldsymbol \xi}}{t_c} + \xi_0 \sqrt{\frac{2}{t_c}} \boldsymbol{\mu}.
      \label{eq:eq_set}
\end{aligned}
\raisetag{-10\baselineskip}
\end{equation}

Here, $\omega_c=\dfrac{q|B|}{m}$ and $\Gamma=\dfrac{\gamma}{m}$. An equivalent matrix representation of the above equations is given by 
\begin{equation}
\dot{\boldsymbol{\chi}} = M \boldsymbol{\chi} + N \boldsymbol{\zeta'},
\label{eq:dynamics-matrix}
\end{equation}
where the matrices $\boldsymbol{\chi}$, $M$, $N$ and $\boldsymbol{\zeta'}$ are given by
\begin{equation}
    \boldsymbol{\chi} = \begin{pmatrix}
        x & y & v_x & v_y & \xi_x & \xi_y
    \end{pmatrix}^{T},
\end{equation}
\begin{equation}
    M = \begin{pmatrix}
        0 & 0 & 1 & 0 & 0 & 0 \\
        0 & 0 & 0 & 1 & 0 & 0 \\
        -\omega_0^2 & \frac{-\alpha+\varepsilon}{m} & -\Gamma & \omega_c & \frac{1}{m} & 0 \\ 
        \frac{-\alpha-\varepsilon}{m} & -\omega_0^2 & -\omega_c & -\Gamma  & 0 & \frac{1}{m} \\
        0 & 0 & 0 & 0 & \frac{-1}{t_c} & 0 \\
        0 & 0 & 0 & 0 & 0 & \frac{-1}{t_c}
    \end{pmatrix},
\end{equation}

\begin{equation}
    N = \begin{pmatrix}
         0 & 0 & 0 & 0 & 0 & 0 \\
         0 & 0 & 0 & 0 & 0 & 0 \\
         0 & 0 &\sqrt{\frac{2 \Gamma T_c}{m}} & 0 & 0 & 0 \\
         0 & 0 & 0 & \sqrt{\frac{2 \Gamma T_h}{m}}  & 0 & 0 \\
         0 & 0 & 0 & 0 &\sqrt{\frac{2 \xi_0^2}{t_c}}  & 0 \\
         0 & 0 & 0 & 0 & 0 & \sqrt{\frac{2 \xi_0^2}{t_c}} \\
    \end{pmatrix},
\end{equation}
and 
\begin{equation}
    \boldsymbol{\zeta'} = (0\ 0\ \zeta_x\ \zeta_y\ \ \mu_x\ \ \mu_y)^T.
\end{equation}

In order to investigate the steady-state properties of the system, we introduce the correlation matrix $\boldsymbol{\Xi}$, with elements defined by
\begin{equation}
    \Xi_{i,j} = \langle \chi_i \chi_j \rangle - \langle \chi_i \rangle \langle \chi_j \rangle.
    \label{eq:corr_mat_def}
\end{equation}

Following the correlation-matrix formalism, the matrix $\boldsymbol{\Xi} $ satisfies the Lyapunov equation \cite{van2007stochastic, gardiner2009stochastic}
\begin{equation}
   M\cdot \boldsymbol{\Xi} + \boldsymbol{\Xi} \cdot M^T + N N^T = 0.
    \label{eq:Xi_relation}
\end{equation}
The solution of the above Eq.~\eqref{eq:Xi_relation} yields the correlation matrix $\boldsymbol{\Xi}$, from which all the physical quantities defined in Eqs.~\eqref{eq:EPR_main}–\eqref{eq:dQh}, \eqref{eq:power_main} and \eqref{eq:Eff_main}  can be obtained~\cite{Adersh2025heatpump}. Since the TUR establishes a fundamental connection between power fluctuations and dissipation, an accurate determination of the entropy production rate is crucial for understanding the precision and performance of the system. Using Eq.~\eqref{eq:EPR_main},
the entropy production rate of the engine can be calculated as
\begin{equation}  
\sigma
=\frac{S}{2}\biggr[\frac{\alpha -\varepsilon +2 \Omega }{T_c}-\frac{\alpha +\varepsilon +2 \Omega }{T_h}\biggr],
\label{eq:EPR}
\end{equation}
where $S$ is given by
\begin{widetext}
\begin{equation}  
\begin{split}
S&=\langle xv_y-yv_x\rangle\\
&=\frac{2 \varepsilon  m^2 \xi _0^2 t_c \omega _c}{G}\Bigr[t_c^2 \left(\Gamma  m^3 \omega _0^2 \omega _c^2-\varepsilon  \omega _c \left(\Gamma ^2 m^2+m^2 \omega _0^2\right)+\Gamma ^3 m^3 \omega _0^2+\alpha ^2 \Gamma  m\right)-m^2 \left(\varepsilon  \omega _c-2 \Gamma  \varepsilon  m^2 t_c \omega _c\right)\Bigr]\\
&+\frac{\Gamma \varepsilon \Delta_1  }{G}\Bigr[\alpha  \Gamma ^2 m^2 \left(T_c-T_h\right)+\varepsilon  \left(T_c+T_h\right) \left(m^2 \omega _c^2+\Gamma ^2 m^2\right)\Bigr]+\frac{2 \Gamma ^2 \varepsilon ^2 \xi _0^2 t_c}{G}\left(m^2 \omega _0^2 t_c^2+\left(\Gamma  t_c+m\right){}^2\right),
\label{eq:S}
\end{split}
    \end{equation}
with
\begin{equation}
G=\Delta_1\bigr[\left(m^2 \omega_c^2+\Gamma^2 m^2\right) \left(\Gamma  \varepsilon  m \omega_c-\varepsilon^2+\Gamma ^2 m^2 \omega_0^2\right)
-\alpha^2 \Gamma^2 m^2\bigr],
\label{eq:G}
\end{equation}
\begin{equation}
\text{and} \quad\Delta_1=\left(m \omega _0^2 t_c^2+\Gamma m  t_c+m\right){}^2+\left(\varepsilon ^2-\alpha ^2\right) t_c^4.
\end{equation}
The parameter $\Omega$ is given by
 \begin{equation}
\Omega=\frac{2 \omega _c^2\left[2 \alpha  \varepsilon  m \xi _0^2 t_c \left(2 \Gamma  t_c+\omega _0^2 t_c^2+1\right)+\omega _0^2 t_c^2 \left(\varepsilon -\Gamma  m \omega _c\right)+2 \Gamma  \varepsilon  t_c+\Delta _2+\varepsilon \right]}{m^2 \omega _c\left[\Gamma  \Delta _1 \varepsilon  \omega _c \left(T_c-T_h\right)-\Gamma  m^2 \xi _0^2 \omega _0^2 t_c^3 \left(\omega _c^2+\Gamma ^2\right)+\varepsilon  m \xi _0^2 t_c \omega _c \left(\left(\Gamma  t_c+1\right){}^2+\omega _0^2 t_c^2\right)-\alpha ^2 \Gamma  \xi _0^2 t_c^3\right]+G_1},
\label{eq:lambda}
\end{equation}
with $G_1$ and $\Delta_2$ as
\begin{equation}
   G_1= m \Gamma ^2\left[\Gamma  \Delta _1 m (\alpha +\varepsilon ) \left(T_c-T_h\right)+\varepsilon  \xi_0^2 t_c \left(m^2 \omega _0^2 t_c^2+\left(\Gamma  m t_c+m\right){}^2\right)\right],
\end{equation}
\begin{equation}
  \text{and} \quad  \Delta _2=\Gamma  \Delta _1 \left[\left(T_c-T_h\right) \left(\varepsilon ^2-\Gamma ^2 m^2 \omega _0^2\right)+\alpha  \varepsilon  \left(T_c+T_h\right)\right].
\end{equation}

\end{widetext}
Now, using Eq.~\eqref{eq:Heat_avg_main}, the heat extracted from the hot and cold baths, $Q_h$ and $Q_c$ can be calculated \cite{Adersh2025heatpump} as
\begin{equation} Q_h=S\biggr[\Omega+\frac{\alpha+\varepsilon}{2}\biggr],
    \label{eq:Qh_OD_main}
\end{equation}
and
\begin{equation}
    Q_c=-S\biggr[\Omega+\frac{\alpha-\varepsilon}{2}\biggr].
\label{eq:Qc_OD_main}
\end{equation}
 
Using Eq.~\eqref{eq:power_main}, the steady state average power calculated as \cite{Adersh2025heatpump}
\begin{equation}  
P=\varepsilon S.
\label{eq:Power_FE_OD}
\end{equation}

It is to be noted that the stability of the system requires that the correlation matrix $\boldsymbol{\Xi}$ be positive definite. In the present model, this condition is met for the parameter $G \ge 0$.
Now, using Eq.~\eqref{eq:Eff_main}, the efficiency $\eta$ ~\cite{Adersh2025heatpump} can be calculated as
\begin{equation}
    \eta=\dfrac{2 \varepsilon }{  \varepsilon+ \alpha +2 \Omega}.
    \label{eq:Eff_1st}
\end{equation}

From Eq.~\eqref{eq:Eff_1st}, it is evident that the efficiency can be $100\% $ ($\eta=1$) for a crictical value of $\Omega$ and this can be obtained by imposing the condition $\eta=1$ in Eq.~\eqref{eq:Eff_1st} and solving for $\Omega$ ($\Omega_{u}$). This $\Omega_{u}$ can be obtained as
\begin{equation}
\Omega_{u} = \frac{\varepsilon - \alpha}{2}.
\label{eq:lambdau}
\end{equation}

Similarly, imposing the Carnot-efficiency condition
\begin{equation}
\eta_c = 1 - \frac{T_{c}}{T_{h}},
\end{equation}
yields the value of $\Omega$ required to reach the Carnot limit. This value, denoted by $\Omega_{ce}$ is given by
\begin{equation}
\Omega_{\mathrm{ce}} = -\frac{\alpha}{2} - \frac{\varepsilon (T_{c}+T_{h})}{2 (T_{c}-T_{h})}.
\label{eq:lambda_ce}
\end{equation}
We obtain two critical values of $\Omega$: $\Omega_{\mathrm{ce}}$ [Eq.~\eqref{eq:lambda_ce}] is the value of $\Omega$ at which the system approaches Carnot efficiency, and $\Omega_{u}$ [Eq.~\eqref{eq:lambdau}] corresponds to the value $\Omega$ at which the system achieves $100\%$ efficiency. Moreover, it should be noted that the desirable condition for the operation of the system for the range of $\Omega$ values is between $\Omega_{\mathrm{ce}}$ and $\Omega_{u}$, as it enables the operation of the system at efficiency that exceed the Carnot limit. Next, the physical quantity of interest is the power fluctuation. Using Eq.~\eqref{eq:dw_main} the power fluctuation $D_P$ can be calculated as
\begin{equation}
\begin{split}
D_P =& \lim_{t \to \infty} 
\langle \left( P(t) -  P  \right)^2\rangle t\\
=& \frac{\varepsilon^2}{2}\left[ (A+A_x)(B+A_y) +A_xA_y+A_{xy}\frac{P\Omega}{\varepsilon \omega_c}\right].
\end{split}
\end{equation}
where the expression for $A, B, A_x, A_y$ and $A_{xy}$ are already given in Appendix A. Now, substituing the value of $D_{p}$ in Eq.~\eqref{eq:TUR_Jtilde}, $\tilde{\bf J}$ can be calculated as 
\begin{equation}   \tilde{\bf J}=\frac{\varepsilon^2}{P^2}\left[ (A+A_x)(B+A_y) +A_xA_y+A_{xy}\frac{P\Omega}{\varepsilon \omega_c}\right].
\label{eq:Jtilde_Full}
\end{equation}
\begin{figure}
\centering
\includegraphics[width=0.99\linewidth]{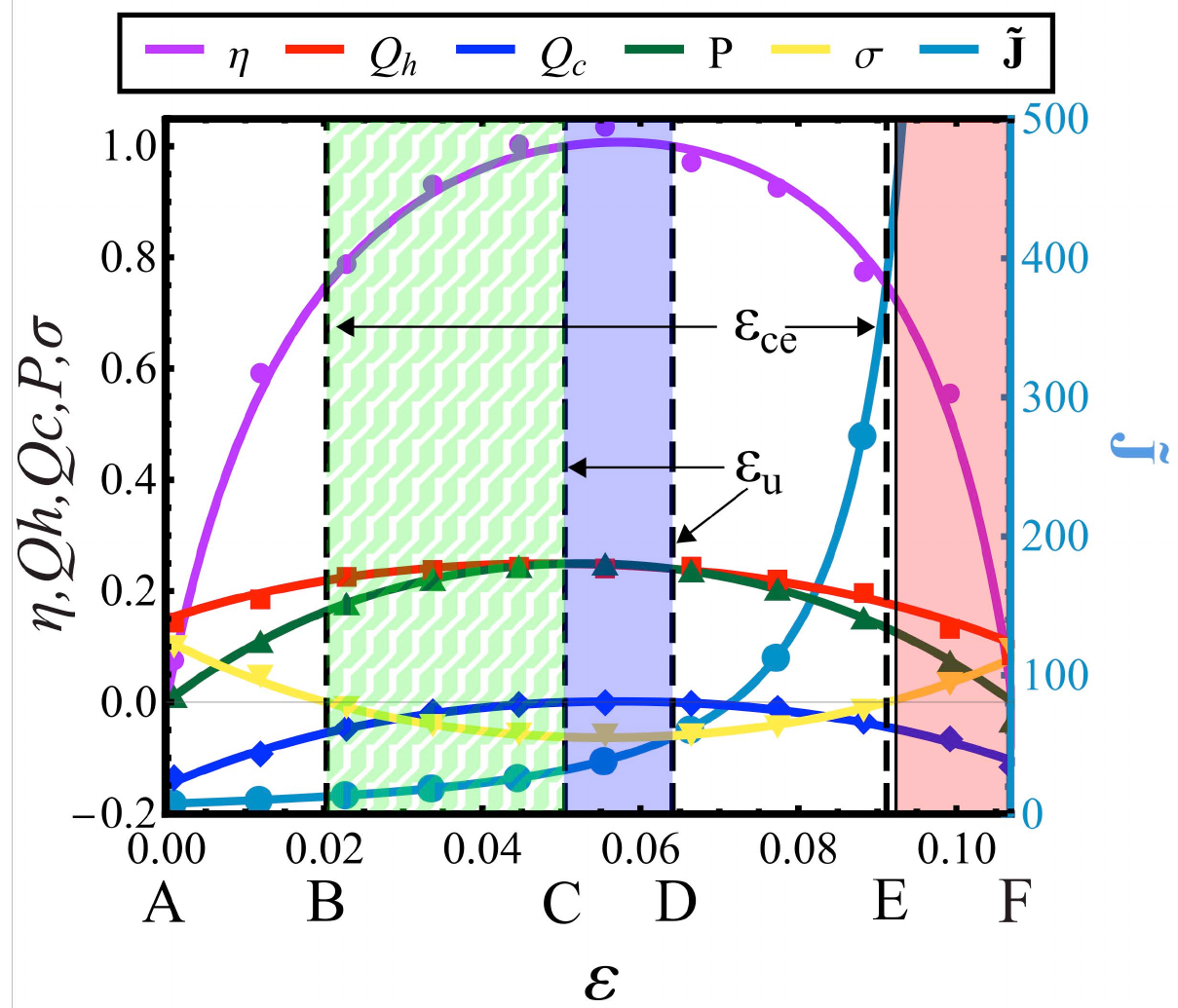}
    \caption{Left axis: The $P$, $Q_h$, $Q_c$, $\eta$ and $\sigma$ as a function of  $\varepsilon$. Right axis: $\tilde{\bf J}$ as a function of  $\varepsilon$. Other common parameters are $t_c =3.0$, $\Gamma=0.1$, $T_c=1.0$, $T_h=4.0$, $\alpha=0.9$, and $m = \xi_0 = \omega_0 = \omega_c=1$. The solid line represents the analytical results, and the dots represent the results obtained from numerical simulation. 
}
\label{fig:TUR_main}
\end{figure}
In order to examine the TUR, we inspect Eq.~\eqref{eq:TUR_main} and it can be observed that TUR can be violated for $\sigma \le 0$. This is because $\tilde{\bf J}$ remains positive definite as power fluctuations remain non-negative and the Boltzmann constant $k_{\mathrm{B}}$ is fundamentally positive. Additionally, the TUR can also be violated when $\sigma$ and $\tilde{\bf J}$ are very low, and their product is near zero. Moreover, from Eq.~\eqref{eq:EPR-withEff} it follows that $\sigma$ becomes negative only for efficiencies exceeding the Carnot bound, i.e. for $\eta > \eta_c$. However, the system achieves Carnot efficiency for the condition $\Omega=\Omega_{ce}$, as confirmed from Eq.~\eqref{eq:lambda}, while for operation beyond the Carnot bound, i.e., for $\eta > \eta_{c}$, the required condition is $\Omega < \Omega_{ce}$. Similarly, for achievement of 100\% efficiency which corresponds to $\eta=1.0$, the satisfied condition is $\Omega=\Omega_{u}$ as confirmed from Eq.~\eqref{eq:lambdau}. Moreover, it is to be noted that the system can even attain  efficiencies greater than unity efficiency ($\eta>1.0$) under the condition $\Omega < \Omega_ {u}$. But exceeding unit efficiency inherently necessitates that both $Q_h$ and $Q_c$ to be positive, which implies simultaneous extraction of heat from both hot and cold baths, resulting the system not operating as an engine. Therefore, unit efficiency ($\eta = 1.0$) represents the upper bound for the operation of system in the engine mode. Whereas, the desirable condition for the system to operate as engine is for values of the parameter $\Omega$ in the range $\Omega_{u} \le \Omega <\Omega_{\mathrm{ce}}$, as it enables  the system in the engine mode to operate at efficiencies exceeding the Carnot limit, but without exceeding the unit efficiency. However, for a fixed $\alpha$, both the crictical values of $\Omega$ ($\Omega_u$ and $\Omega_{ce}$) depends on the load strength $\varepsilon$. One can compute the $\varepsilon$ values corresponding to both $\Omega_{ce}$ and $\Omega_u$ ($\varepsilon_{ce}$ and $\varepsilon_u$) by substituting Eq.~\eqref{eq:lambda_ce} and Eq.~\eqref{eq:lambdau} in Eq.~\eqref{eq:lambda}. The resulting equation becomes a quartic equation with two real solutions for both $\varepsilon_{ce}$ and $\varepsilon_{u}$ corresponding to the $\varepsilon$ values at which $\eta$ becomes $\eta_c$ and $\eta$ becomes $1.0$.
\begin{figure*}
\centering
\includegraphics[width=0.98\linewidth]{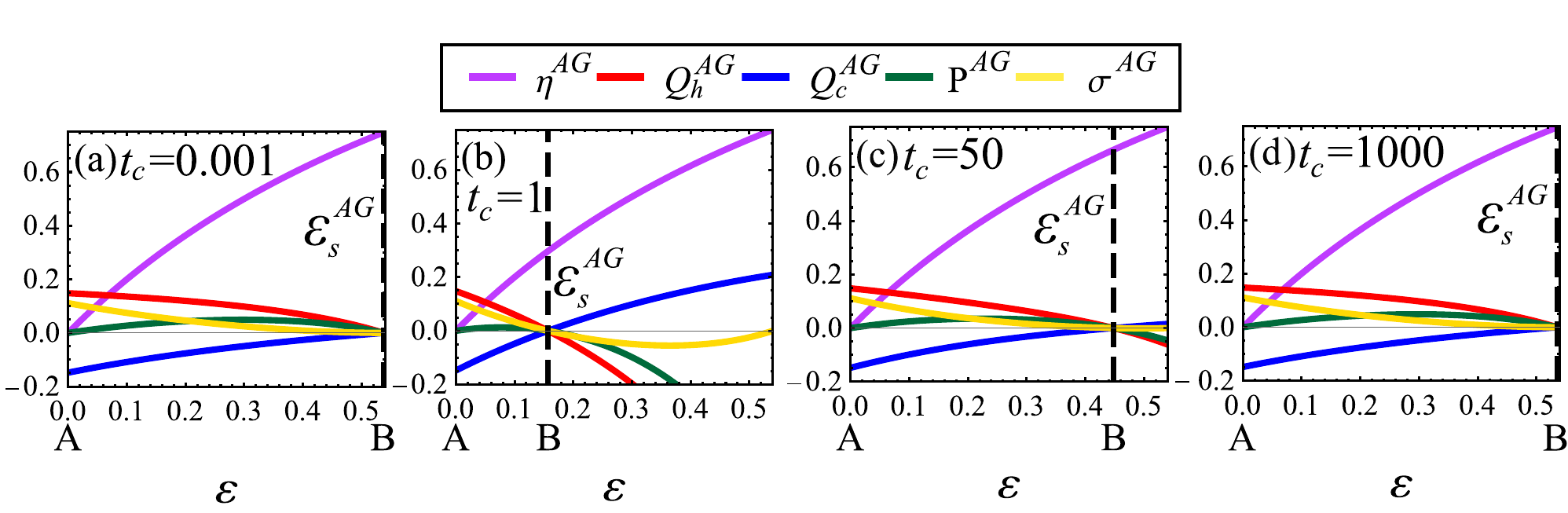}
    \caption{The evolution of $P^{AG} , Q^{AG}_h, Q^{AG}_c, \eta^{AG} ~\text{and}~ \sigma^{AG}$ as a function $\varepsilon$ is plotted in (a) for $t_c=0.001$, in (b) for $t_c=1.0$, in (c) for $t_c=50.0$,
    and in (d) for $t_c=1000.0$. The other common parameters are $\Gamma=0.1$, $T_c=1.0$, $T_h=4.0$, $\alpha=0.9$, and $m = \xi_0 = \omega_0 =1.0$.}
    \label{fig:Fig.2}
\end{figure*}

To understand the dependence of all physical parameters on $\varepsilon$, in Fig.~\ref{fig:TUR_main}, we have shown the evolution of $P$, $Q_h$, $Q_c$, $\eta$, $\sigma$, and ${\bf \tilde{J}}$ as a function of $\varepsilon$. From the figure, it is evident that the quantities $P$, $Q_h$, $Q_c$, and $\eta$ initially increase with increasing $\varepsilon$, approach a maximum value and subsequently decrease to a minimum value. In contrast, we observe that $\sigma$ initially decreases as a function of $\varepsilon$, approaches a minimum value, and then subsequently increases  to a maximum value, whereas $\bf \tilde{J}$ increases monotonically with load strength. Since both $P$ and $\eta$ exhibit a similar behavior as a function of $\varepsilon$, with both the maximum power and the maximum efficiency occurring at the same $\varepsilon$ value, the conventional trade-off between power and efficiency is violated. Additionally, we observe that $P$, $Q_h$, and $\eta$ are positive, while $Q_c$ is negative, for the values of $\varepsilon$ corresponding to the ranges from $A$ to $C$ and from $D$ to $F$, which confirms the operation of the system in the engine mode. Moreover, for lower values of $\varepsilon$, i.e., for $\varepsilon$ values ranging from $A$ to $B$, $Q_h$ remains positive while $Q_c$ remains negative. As $P=(Q_h+Q_c)$ and $Q_c$ is negative in this range of $\varepsilon$, the power $P$ satisfies the condition $P<Q_h$ throughout this range, and hence the efficiency remains less than unity.

However, with increasing values of $\varepsilon$, both the values of $P$ and $Q_h$ increase and at the same time the magnitude of $Q_{c}$ decreases towards zero. This results in the increase in the value of $\eta$ and exactly at $\varepsilon=\varepsilon_{ce}$ (i.e, at point B), the system achieves the Carnot efficiency $\eta_{c}$. Upon further increase in $\varepsilon$ values, the magnitude of $Q_c$ decreases and eventually vanishes at $\varepsilon=\varepsilon_{u}$ corresponding to the point $C$. At this point, the output power becomes equal to the heat absorbed from the hot bath, i.e., $P = Q_h$. This corresponds to the situation at which all the heat extracted from the hot bath is converted into work without any energy transfer to the cold bath and hence the system approaches an efficiency of $100\%$. Beyond this point $C$, with a further increase in the values of $\varepsilon$, $Q_c$ becomes positive and hence $P>Q_{h}$. This results for the system attaining efficiency greater than unity. However, both positive $Q_h$ and $Q_c$ imply that the system cannot be operationally classified as an engine as it corresponds to the simultaneous extraction of heat from both the hot and cold baths. With a further increase in $\varepsilon$ value, $Q_c$ decreases and eventually vanishes at $\varepsilon=\varepsilon_{u}$ corresponding to the point $D$ and $\eta$ becomes unity again. Beyond this point $D$, an increase in $\varepsilon$ value causes $Q_c$ to be negative again and the magnitude of $Q_c$ increases with increasing $\varepsilon$. Consequently, the efficiency decreases from unity and reaches the Carnot efficiency at $\varepsilon=\varepsilon_{ce}$ corresponding to point $E$. Beyond this point E, with further increase in $\varepsilon$ values, the efficiency $\eta$ decreases. Although Carnot efficiency can be achieved at two distinct values of $\varepsilon$ at the points $B$ and $E$, the corresponding value of fluctuation ${\bf \tilde{J}}$ is substantially higher at the point $E$. Similarly, although the system achieves $100\%$ efficiency twice (at points $C$ and $D$) throughout the whole range of $\varepsilon$, the corresponding ${\bf \tilde{J}}$ is higher at the point $D$. Therefore, the system in the $\varepsilon$ range from $B$ to $C$ performs as a more operationally favorable engine, as it enables the achievement of Carnot efficiency ($\eta=\eta_c$) and $100\%$ efficiency ($\eta=1.0$) while maintaining minimal power fluctuations.  

 Additionally, it is interesting to note that throughout this range of $\varepsilon$ values  from $B$ to $C$, the system achieves Carnot efficiency and even $100\%$ efficiency, maintaining with zero and negative entropy production, however without blowing off the power fluctuations. This is a clear confirmation of the violation of TUR [Eq.\eqref{eq:TUR_main}] \cite{Pietzonka2018universal}. This is because the zero or negative value of $\sigma$ does not validate the inequality of TUR [Eq.\eqref{eq:TUR_main}] as $k_{B}$ is the Boltzmann constant, which is finite positive \cite{arghya2025violation}. Notably, the TUR is also violated for $\varepsilon$ values from $A$ to $B$, due to the low values of $\sigma$ and $\tilde{\bf J}$, despite not achieving Carnot or unit efficiency.
 
Next, in the absence of a magnetic field (i.e., in the $\omega_c \to 0$ limit), we analyze the physical parameters for the performance characteristics of the system. In this limit, system becomes an active gyrator and hence the dynamics of the particle described by Eq.~\eqref{eq:maineqmotion-vector} reduces to that of an active gyrator and can be rewritten as
\begin{equation} 
m\ddot{{\bf r}}(t) =-\gamma \dot{{\bf r}}(t)- \nabla V +{\bf F}^{nc}+{{\boldsymbol{\xi}}}(t)+ {{\boldsymbol{\zeta}}}(t).
\label{eq:maineqmotion-vector_AG}
\end{equation}
Using Eq.~\eqref{eq:power_main}
the corresponding output power $P^{AG}$ can be obtained~\cite{Adersh2025heatpump} and is given by
\begin{align}
P^{AG}&=\frac{\varepsilon}{G_{A}}\biggl\{ \Gamma \varepsilon \Delta _1 m^2\left(\alpha \Gamma ^2 \left(T_c-T_h\right)+\Gamma ^2 \varepsilon  \left(T_c+T_h\right)\right)\nonumber\\
&+2 \Gamma ^2 \varepsilon ^2 t_c \xi _0^2\left(m^2 \omega _0^2 t_c^2+\left(\Gamma  t_c+m\right)^2\right)\biggl\},
\label{eq:S_AG}
\end{align}
\begin{equation}
\text{with}~~G_A= \Delta _1\left(\Gamma ^2 m^2 \left(\Gamma ^2 m^2 \omega _0^2-\varepsilon ^2\right)-\alpha ^2 \Gamma ^2 m^2\right).\qquad\quad\nonumber
\label{eq:G_AG}
\end{equation}
Now, using Eq.~\eqref{eq:Heat_avg_main}, the expressions for $Q_h^{AG}$ and $Q_c^{AG}$  can be calculated as \cite{Adersh2025heatpump}
\begin{align}  Q^{AG}_h&=\biggr[\frac{\alpha+\varepsilon}{2G_A}\biggr]\biggl\{ \Gamma \varepsilon \Delta _1 m^2\left(\alpha \Gamma ^2 \left(T_c-T_h\right)+\Gamma ^2 \varepsilon  \left(T_c+T_h\right)\right)\nonumber\\
    &+2 \Gamma ^2 \varepsilon ^2 t_c \xi _0^2\left(m^2 \omega _0^2 t_c^2+\left(\Gamma  t_c+m\right){}^2\right)\biggl\},
\end{align}
and
\begin{align}
Q^{AG}_c&=\biggr[\frac{\varepsilon-\alpha}{2G_A}\biggr]\biggl\{ \Gamma \varepsilon \Delta _1 m^2\left(\alpha \Gamma ^2 \left(T_c-T_h\right)+\Gamma ^2 \varepsilon  \left(T_c+T_h\right)\right)\nonumber\\
&+2 \Gamma ^2 \varepsilon ^2 t_c \xi _0^2\left(m^2 \omega _0^2 t_c^2+\left(\Gamma  t_c+m\right)^2\right)\biggl\}. 
\label{eq:Qc:AG}
\end{align}
Further, using Eq.~\eqref{eq:Eff_main}, the efficiency of the active gyrator $\eta^{AG}$ ~\cite{Adersh2025heatpump} can be calculated as
\begin{equation}
\eta^{AG}=\frac{2 \varepsilon}{\alpha+\varepsilon}.
\label{eq:eta_ag}
\end{equation}
 From the above equation, it is confirmed that $\eta_{AG}$ is independent of $t_c$. Similarly, using Eq.~\eqref{eq:EPR_main},
the entropy production rate of the engine can be calculated as
\begin{align}
 \sigma^{AG}&=-\left(\frac{\Gamma ^2 m^2 \varepsilon \left((\alpha +\varepsilon ) T_c+(\varepsilon -\alpha ) T_h\right)}{2 G_A T_c T_h}\right)\biggl[4 \Gamma  \varepsilon  m \xi _0^2 t_c^2\nonumber\\
 &+2 \Gamma ^2 \varepsilon  \xi _0^2 t_c^3+m^2 \Bigl(2 \varepsilon  \xi _0^2 \omega_0^2 t_c^3+2 \varepsilon  \xi _0^2 t_c+\Gamma  \Delta _1 (\alpha +\varepsilon ) T_c\nonumber\\
 &+\Gamma \Delta _1 (\varepsilon -\alpha ) T_h\Bigr)\biggr].
\end{align}

In Fig.~\ref{fig:Fig.2}, we show the evolution of $P^{AG}$, $Q^{AG}_h$, $Q^{AG}_c$, $\eta^{AG}$ and $\sigma^{AG}$ as a function of  $\varepsilon$ for different values of $t_c$  for an active Brownian gyrator. From Fig.~\ref{fig:Fig.2}(a), it is observed that for the values of $\varepsilon$ in the range $A$ to $B$, $P^{AG}$, $Q^{AG}_{h}$ and $\eta^{AG}$ remain positive, while $Q^{AG}_{c}$ remains negative. Hence, in this range, the system operates in the engine mode. In addition, it is observed that $P^{AG}$ exhibits a non-monotonic dependence on $\varepsilon$, while $Q^{AG}_h$ decreases monotonically with $\varepsilon$. In contrast, $\eta^{AG}$ increases with increasing $\varepsilon$ values. Furthermore, the engine stalls at $\varepsilon=\varepsilon^{AG}_s$ (i.e, at point $B$), where the output power vanishes ($P^{AG}=0$). Since $\eta^{AG}$ increases monotonically with $\varepsilon$, its maximum value is attained in the vicinity of the stalling point. Beyond the stalling point $B$, the system does not operate as a heat engine as $P^{AG}$ becomes less than zero. 
With an additional increase in $t_{c}$ values [see Figs.~\ref{fig:Fig.2}(b), \ref{fig:Fig.2}(c) and \ref{fig:Fig.2}(d)], similar behavior is observed for $P^{AG}, Q^{AG}_h$, $\eta^{AG}$ and $Q^{AG}_{c}$, while the engine is observed to stall faster initially and then gets delayed with activity. As a result, the stalling point varies non-monotonically with activity. Specifically, $\varepsilon^{AG}_s$ decreases with increasing $t_c$, attains a minimum, then increases with increase in $t_c$ value, and eventually saturates for large $t_c$ values. Moreover, it is observed that even though the system attains maximum efficiency in the vicinity of stalling, the Carnot efficiency can't be achieved in case of an active gyrator. It is also observed that $\sigma^{AG}$ shows a decreasing behavior with $\varepsilon$ and vanishes when the engine stalls. Although a power-efficiency trade-off is evident in this case, it differs from the conventional power-efficiency trade-off~\cite{Pietzonka2018universal,Lee2020Brownian}, where the output power vanishes as the efficiency approaches the Carnot limit.

However, in the absence of activity (i.e., in $t_c \to 0$ limit), the dynamics described by Eq.~\eqref{eq:maineqmotion-vector} reduces to that case of a Brownian magneto gyrator and can be rewritten as
\begin{equation} 
m\ddot{{\bf r}}(t) =-\gamma \dot{{\bf r}}(t)+|q|\left({{\bf{\dot{r}}}(t)\times{\bf B}}\right) - \nabla V +{\bf F}^{nc}+{{\boldsymbol{\zeta}}}(t).
\label{eq:maineqmotion-vector_MG}
\end{equation}

In this case, the expression of output power $P^{MG}$ [Eq.~\eqref{eq:power_main}] can be calculated as \cite{Adersh2025heatpump}  
\begin{equation}
    P^{MG}=\varepsilon S_1,
    \label{eq:P_MG}
\end{equation}
with $S_1$ as
\begin{equation}
    S_1=\frac{\Gamma  \varepsilon \left[\alpha  \Gamma ^2 m^2 \left(T_c-T_h\right)+\varepsilon  \left(T_c+T_h\right) \left(m^2 \omega _c^2+\Gamma ^2 m^2\right)\right]}{\left(m^2 \omega _c^2+\Gamma ^2 m^2\right) \left(\Gamma  \varepsilon  m \omega _c-\varepsilon ^2+\Gamma ^2 m^2 \omega_0^2\right)-\alpha ^2 \Gamma ^2 m^2}.
    \label{eq:S_MG}
\end{equation}
\begin{widetext}
\hspace{-0.35cm}Now, using Eq.~\eqref{eq:Heat_avg_main}, the expressions for $Q^{MG}_h$ and $Q^{MG}_c$  can be calculated as \cite{Adersh2025heatpump}
\begin{equation}
Q^{MG}_h=S_1\left[\frac{2 \omega _c^2 \left(\Gamma  m^2 \left(\left(T_c-T_h\right) \left(\varepsilon ^2-\Gamma ^2 m^2 \omega _0^2\right)+\alpha  \varepsilon  \left(T_c+T_h\right)\right)+\varepsilon \right)}{\Gamma ^3 m^4 (\alpha +\varepsilon ) \left(T_c-T_h\right)+\Gamma  \varepsilon  m^4 \omega _c^2 \left(T_c-T_h\right)}+\frac{\alpha +\varepsilon }{2}\right],
    \label{eq:Qh_MG}
\end{equation}
and
\begin{equation}
Q^{MG}_c=-S_1\left[\frac{2 \omega _c^2 \left(\Gamma  m^2 \left(\left(T_c-T_h\right) \left(\varepsilon ^2-\Gamma ^2 m^2 \omega _0^2\right)+\alpha  \varepsilon  \left(T_c+T_h\right)\right)+\varepsilon \right)}{\Gamma ^3 m^4 (\alpha +\varepsilon ) \left(T_c-T_h\right)+\Gamma  \varepsilon  m^4 \omega _c^2 \left(T_c-T_h\right)}+\frac{\alpha -\varepsilon }{2}\right].
\label{eq:Qc_MG}
\end{equation}
Using Eq.~\eqref{eq:Eff_main}, the efficiency of the Brownian magneto gyrator $\eta^{MG}$ can be calculated as
\begin{equation}
\eta^{MG}=\frac{2 \Gamma  \varepsilon  m^4 \left(T_c-T_h\right) \left(\Gamma ^2 (\alpha +\varepsilon )+\varepsilon  \omega _c^2\right)}{4 \varepsilon  \omega _c^2+\Gamma  m^4 \left(T_c-T_h\right) \left[\Gamma ^2 (\alpha +\varepsilon )^2+\omega _c^2 \left(\varepsilon  (\alpha +\varepsilon )-4 \Gamma ^2 \omega _0^2\right)\right]+4 \Gamma  \varepsilon  m^2 \omega _c^2 \left[(\alpha +\varepsilon ) T_c+(\alpha -\varepsilon ) T_h\right]}.
    \label{eq:Eff_mg}
\end{equation}
\end{widetext}
From the above equation, it is observed that $\eta^{MG}$ has a dependence on $\omega_{c}$, however in this case of magneto gyrator Carnot efficiency can not be achieved.
\begin{figure}
\includegraphics[width=1.0\linewidth]{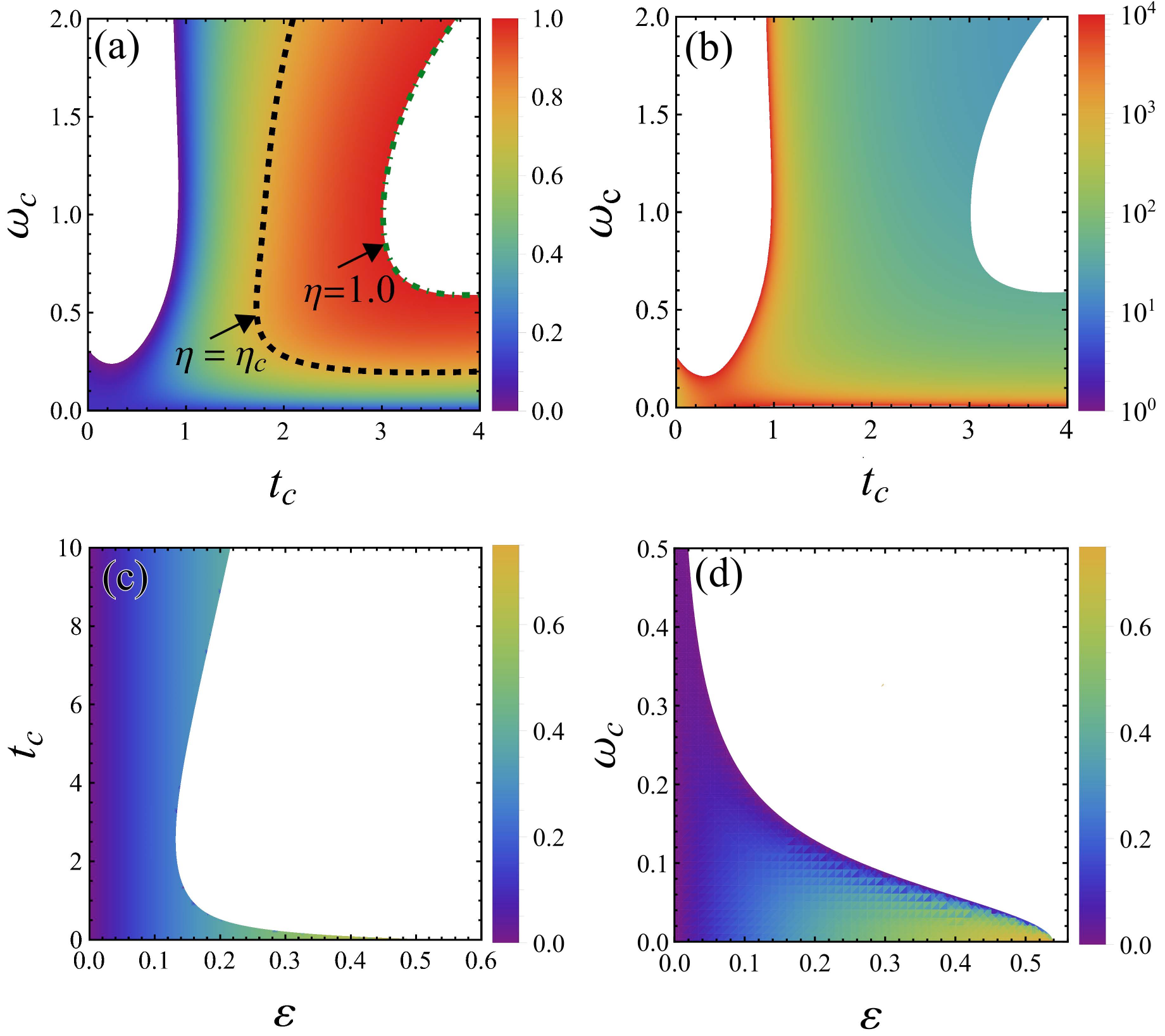}
\caption{(a) The 2D parametric plot of $\eta$ as a function of $\omega_c$ and $t_c$ for $\varepsilon=0.05$ and (b) the corresponding $\mathbf{\tilde{J}}$ as a function of $\omega_c$ and $t_c$. (c) The 2D parametric plot of $\eta^{AG}$ as a function of $t_c$ and $\varepsilon$  and (d) the 2D parametric plot of $\eta^{MG}$ as a function of $\omega_c$ and $\varepsilon$. The other common parameters are $\Gamma=0.1$, $T_c=1.0$, $T_h=4.0$, $\alpha=0.9$, and $m = \xi_0 = \omega_0 =1.0$.}
\label{fig:Fig3}
\end{figure}
Similarly, in the absence of both activity and magnetic field (i.e., in the limit $t_c \to 0$ and $\omega_c \to 0$), the dynamics described by Eq.~\eqref{eq:maineqmotion-vector} reduces to that of a Brownian gyrator and can be rewritten as
\begin{equation} 
m\ddot{{\bf r}}(t) =-\gamma \dot{{\bf r}}(t)- \nabla V +{\bf F}^{nc}+{{\boldsymbol{\zeta}}}(t).
\label{eq:maineqmotion-vector_BG}
\end{equation}
In this case, the expression for $P^{BG}$ [Eq.~\eqref{eq:power_main}] can be calculated as
\begin{equation}
P^{BG}=\frac{\Gamma  \varepsilon ^2\left[\alpha  \Gamma ^2 m^2 \left(T_c-T_h\right)+\Gamma ^2 \varepsilon  m^2 \left(T_c+T_h\right)\right]}{\Gamma ^4 m^4 \omega _0^2-\Gamma ^2 m^2 \left(\alpha ^2+\varepsilon ^2\right)}.
\label{eq:P:BG}
\end{equation}
Using Eq.~\eqref{eq:Heat_avg_main}, the expressions for $Q^{BG}_h$ and $Q^{BG}_c$  can be obtained as
\begin{equation}
    Q^{BG}_h=-\frac{\Gamma  \varepsilon  (\alpha +\varepsilon )\left[(\alpha +\varepsilon ) T_c+(\varepsilon -\alpha ) T_h\right]}{2 \left(\alpha ^2+\left(\varepsilon +\Gamma  m \omega _0\right) \left(\varepsilon -\Gamma  m \omega _0\right)\right)},
    \label{eq:Qh:BG}
\end{equation}
and
\begin{equation}
       Q^{BG}_c=\frac{\Gamma  \varepsilon  (\alpha -\varepsilon )\left[(\alpha +\varepsilon ) T_c+(\varepsilon -\alpha ) T_h\right]}{2 \left(\alpha ^2+\left(\varepsilon +\Gamma  m \omega _0\right) \left(\varepsilon -\Gamma  m \omega _0\right)\right)}.
    \label{eq:Qc:BG}
\end{equation}
Using Eq.~\eqref{eq:Eff_main}, the efficiency of the Brownian  gyrator $\eta^{BG}$ can be obtained as
\begin{equation}
\eta^{BG}=\frac{2 \varepsilon}{\alpha+\varepsilon},
\label{eq:eta_bg}
\end{equation}
which is same as that case of an active gyrator [Eq.~\eqref{eq:eta_ag}]. It is observed that $\eta^{AG}$($=\eta^{BG}$) is independent of $t_c$.  However, both output power and heat has dependence on $t_{c}$ and are different for both active and Brownian gyrator. The efficiency has dependence on activity time scale  only in the presence of magnetic field. 

In case of Brownian gyrator, the system can achieve Carnot efficiency as a function of $\varepsilon$ and the critical epsilon ($\varepsilon_{ce}^{BG}$), at which $\eta^{BG}$ becomes $\eta_{c}$ can be obtained as
\begin{equation}
\varepsilon^{BG}_{ce}=\frac{\alpha  \left(T_h-T_c\right)}{T_c+T_h}.
\label{eq:epsilonce_BG}
\end{equation}
At this point, the entropy production rate of the engine can be calculated as
\begin{equation}
\sigma^{BG}=\frac{\alpha  \Gamma  \left(T_c-T_h\right) \left[(\alpha +\varepsilon ) T_c+(\varepsilon -\alpha ) T_h\right]}{2 T_c T_h \left(\alpha ^2+\Gamma ^2 \omega_0^2\right)}.
\end{equation}
In order to understand the usability of the engine, it is essential to understand the role of fluctuations in its output power. Therefore, using Eq.~\eqref{eq:TUR_Jtilde}, we calculate $\tilde{\bf J}$ as
\begin{equation}
    {\bf \tilde{J}}=\frac{\varepsilon ^2  \omega _0^2}{2}  \left[\frac{\left(T_c+T_h\right){}^2}{m^2 \omega _0^4-\alpha ^2}-\frac{\Gamma ^4 \left(T_c-T_h\right){}^2}{m^2 \left(\alpha ^2+\Gamma ^2 \omega _0^2\right){}^2}\right].
\end{equation}

Figure~\ref{fig:Fig3} illustrates the evolution of $\eta$ for different gyrator models, for parameter regimes where it operates as an engine. The two dimensional evolution of $\eta$ of the active magneto gyrator as a function of $\omega_{c}$ and $t_{c}$ is shown in Fig.~\ref{fig:Fig3}(a) and the corresponding fluctuations $\tilde{\bf J}$ as a function of $\omega_c$ and $t_c$ is shown in Fig.~\ref{fig:Fig3}(b) for a fixed load strength, $\varepsilon=0.05$. From Fig.~\ref{fig:Fig3}(a), it is evident that the system can achieve Carnot efficiency and even the maximum efficiency can be $100\%$. The dashed lines denote the points where the system reaches Carnot, and the green dots denote the points where the system reaches unit or $100\%$ efficiency. The corresponding $\tilde{\bf J}$ values remain minimal at these points [see Fig.~\ref{fig:Fig3}(b)]. In Fig.~\ref{fig:Fig3}(c), we have shown the two dimensional plot of efficiency of the active gyrator $\eta^{AG}$ as a function of $t_c$ and $\varepsilon$. In Fig.~\ref{fig:Fig3}(d), we show the plot of efficiency of the Brownian magneto gyrator  $\eta^{MG}$ as a function of $\omega_c$ and $\varepsilon$. From Fig.~\ref{fig:Fig3}(c), it is observed that Carnot efficiency cannot be achieved and the $\eta^{AG}$ does not have a dependence on $t_{c}$, which is consistent with Eq.~\eqref{eq:eta_ag}. Similarly, from Fig.~\ref{fig:Fig3}(d), it is observed that for lower values of $\omega_c$, $\eta^{MG}$ increases as a function of $\varepsilon$, whereas it exhibits a non-monotonic dependence on $\varepsilon$ for larger values of $\omega_c$. Moreover, for a fixed $\varepsilon$ value, $\eta^{MG}$ decreases monotonically with increase in value of $\omega_c$. 

\begin{figure}
    \centering   \includegraphics[width=1.0\linewidth]{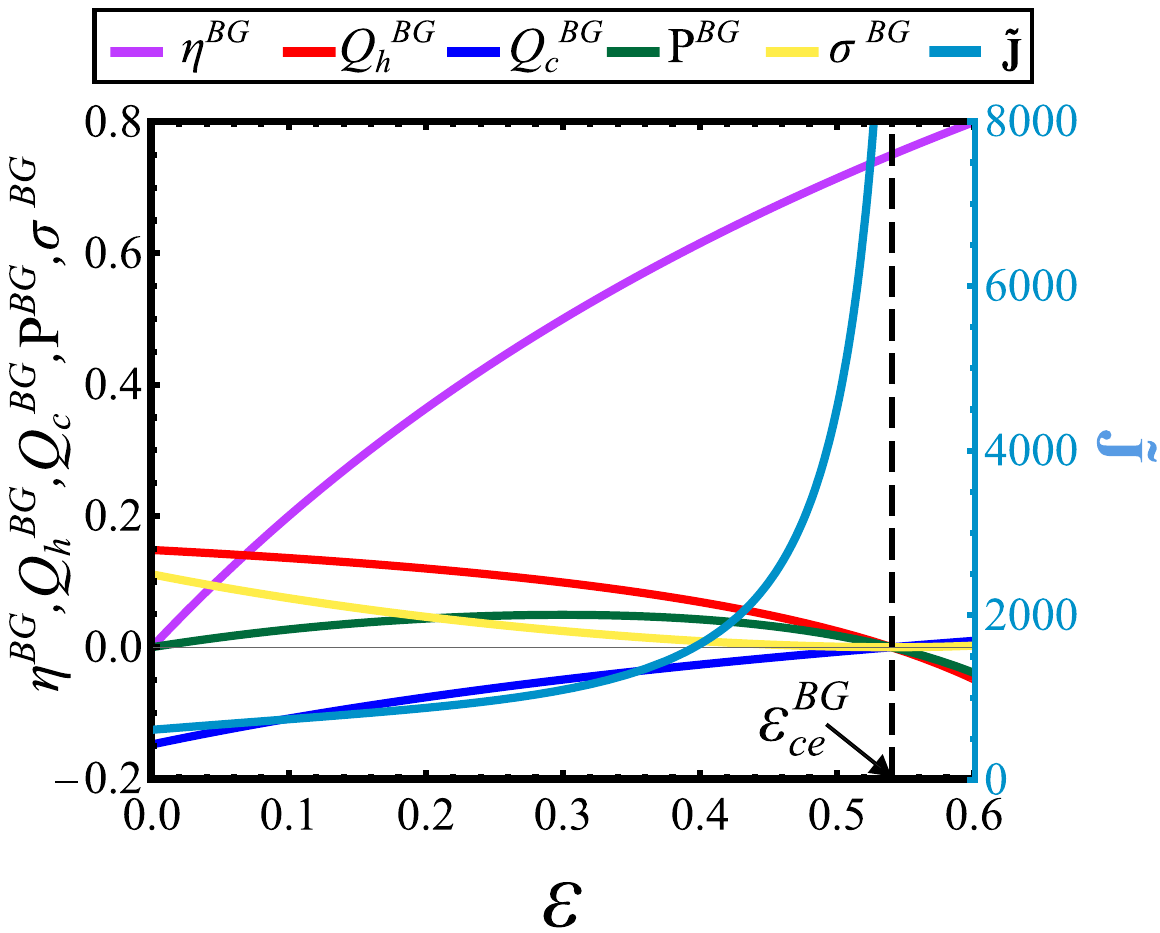}
    \caption{Left axis: The $P^{BG}$, $Q^{BG}_h$, $Q^{BG}_c$, $\eta^{BG}$ and $\sigma^{BG}$ as a function of  $\varepsilon$. Right axis: $\tilde{\bf J}$ as a function of  $\varepsilon$. Other common parameters are  $\Gamma=0.1$, $T_c=1.0$, $T_h=4.0$, $\alpha=0.9$, and $m = \xi_0 = \omega_0 = 1$.}
\label{fig:BG:Full}
\end{figure}

In Fig.~\ref {fig:BG:Full}, we have shown the evolution of $P^{BG}$, $Q^{BG}_h$, $Q^{BG}_c$, $\eta^{BG}$, $\sigma^{BG}$, and ${\bf \tilde{J}}$ as a function of $\varepsilon$ for a Brownian gyrator. From the figure, it is evident that $P^{BG}$ shows a non-monotonic dependence on $\varepsilon$, showing a maximum at an intermediate $\varepsilon$ and then subsequently decreasing to zero as the engine stalls. In contrast, $Q^{BG}_h$ and $\sigma^{BG}$ shows decreasing behavior as functions of $\varepsilon$. 
Furthermore, $\eta^{BG}$ increases with increase in $\varepsilon$ and at exactly $\varepsilon=\varepsilon^{BG}_{ce}$, the system achieves the Carnot efficiency as $P^{BG}$ goes to zero, indicating that the conventional power-efficiency trade-off is satisfied~\cite{Lee2020Brownian}. Additionally, it is observed that $\eta^{BG}$ and $\bf \tilde{J}$ are both increasing functions of $\varepsilon$, which implies that higher efficiencies are accompanied by larger fluctuations. As the entropy production approaches zero, the fluctuations in the output power blow off, illustrating the impossibility of simultaneously achieving low dissipation and highly precise power output, thereby validating the TUR. Hence, the system even though achieves Carnot efficiency, becomes unsuitable for practical applications as the fluctuations blow up. Moreover, $\eta^{BG}=\eta_{c}$ represents the maximum achievable efficiency of the system while operating as a Brownian gyrator.
\section{CONCLUSIONS}\label{sec:summary}
In this work, we have investigated the thermodynamic performance characteristics of an active magneto gyrator while operating in the mode of heat engine. The system consists of a charged active particle confined within a two-dimensional asymmetric parabolic potential and simultaneously coupled to two thermal baths maintained at two different temperatures. In addition, a uniform magnetic field is applied perpendicular to the plane of motion. The combined presence of both potential asymmetry and a temperature gradient results in a sustained gyrating motion of the particle. As a result, the particle exerts a nonzero average torque on the confining potential. By introducing an external load force, we explore the thermodynamic performance characteristics and analyze the fluctuations in the output power.

The power and efficiency of the active magneto gyrator exhibit a nonmonotonic dependence on the load strength. Moreover, in contrast to both the Brownian magneto gyrator and the purely active gyrator, the efficiency of the active magneto gyrator is not subjected to a universal upper bound. Interestingly, by tuning the system parameters, the efficiency can approach the Carnot limit and may even reach unity. Therefore, achieving efficiencies equal to or exceeding the Carnot efficiency is necessarily accompanied by a violation of the TUR, since this bound does not hold in regimes of vanishing or negative entropy production. Moreover, the fluctuations of the output power remain minimal even when the entropy production rate approaches zero or becomes negative. This regime is therefore particularly advantageous for operating as a heat engine, as it allows for finite power output at and beyond the Carnot limit while maintaining low power fluctuations. By contrast, neither the Brownian magneto gyrator nor the purely active gyrator can attain Carnot efficiency while simultaneously achieving non zero power. In the case of the Brownian gyrator, the Carnot efficiency represents the maximum attainable efficiency and is reached only when power fluctuations diverge, thus severely limiting operational precision. Finally, we believe that our model can provide a useful framework for the design of nanoscale engines and synthetic molecular machines~\cite{kay2007synthetic} and serve as a prototypical autonomous microscopic heat engine~\cite{filliger2007brownian}. 
An interesting extension of our present work would be to explore the collective phenomena due to many interacting active magneto gyrators like rotating molecular motors in membranes \cite{peter2003membrane} and rotating magnetic disks in a two-dimensional interfaces \cite{Grzybowski2000Dynamic}. 

\section{Acknowledgement}
We thank Prof. Sriram Ramaswamy for valuable discussions. MS acknowledges the SERB-SURE grant (SUR/2022/000377), CRG grant (CRG/2023/002026) from DST, Govt. of India, start-up grant from the Faculty Recharge program (FRP-56055) of UGC, and the departmental computational facility at the Department of Physics, University of Kerala,  for financial support. 
\begin{widetext}
\appendix
\section{} \label{sec:app_A}
\begin{equation}
\begin{split}
A =& -2\Biggl\{
\Gamma \Delta_1  \Big[
\Gamma \varepsilon m^2 \omega_0^2 \big((\alpha+\varepsilon)T_c + (\alpha-\varepsilon)T_h\big)
+ \Gamma \varepsilon m^2 \omega_c^2 \big((\alpha+\varepsilon)T_c + (\alpha-\varepsilon)T_h\big) - \varepsilon^2 m^2 \omega_c \big((\alpha+\varepsilon)T_c + (\alpha-\varepsilon)T_h\big)
\\ &+ \Gamma^3 m^2 \omega_0^4 (T_h - T_c)
+ \alpha \Gamma^3 \big((\alpha+\varepsilon)T_c + (\varepsilon-\alpha)T_h\big)
\Big]
+ 2\alpha\varepsilon\,\xi_0^2 t_c \Big[
m^4 \Big(
\Gamma \big(\omega_0^4 t_c^2 + 3\omega_0^2 t_c^2 \omega_c^2 + t_c^2\omega_c^4
+ \omega_c^2 + \omega_0^2\big)
\\ &- \varepsilon \omega_c \big(t_c^2(\omega_c^2 + 2\omega_0^2)+1\big)
\Big)+ 2 m^3 t_c \Big(
\Gamma^2 (\omega_0^2 t_c^2 \omega_c^2 + \omega_c^2 + \omega_0^2)
+ \Gamma \varepsilon \omega_c (t_c^2\omega_c^2 - 1)
- \varepsilon^2 t_c^2 \omega_c^2
\Big)+ \Gamma m^2 \Big(
\Gamma^2 + t_c^2\big(\alpha^2 + 3\Gamma^2\omega_0^2 \\ &- (\varepsilon - \Gamma \omega_c)(2\Gamma\omega_c + \varepsilon)\big)
\Big)+ 2\Gamma^2 m t_c \Big(
\Gamma^2 + t_c^2(\alpha^2 + \Gamma^2\omega_0^2 + \Gamma\varepsilon\omega_c - \varepsilon^2)
\Big)
+ \Gamma^5 t_c^2
\Big]
\Biggl\}.
\end{split}
\end{equation}

\begin{equation}
\begin{split}
B &= 2 \Gamma^2 \Delta_1   \Bigl[
\varepsilon T_h (\alpha + \Gamma \omega_c - \varepsilon)
+ \varepsilon T_c (\alpha - \Gamma \omega_c + \varepsilon)
- \Gamma^2 \omega_0^2 T_c
+ \Gamma^2 \omega_0^2 T_h
\Bigr]+ 4 \alpha \varepsilon \xi_0^2 t_c \Bigl[
m^2 \bigl(\Gamma t_c^2 (\omega_c^2 + \omega_0^2) - \varepsilon t_c^2 \omega_c + \Gamma \bigr)
\\
&+ 2 \Gamma^2 m t_c
+ \Gamma^3 t_c^2
\Bigr].
\end{split}
\end{equation}

\begin{equation}
\begin{split}
A_{xy}&=
\Gamma \Delta_1   \Biggl\{
\Gamma\!\left[(\alpha+\varepsilon)T_c+(\alpha-\varepsilon)T_h\right]\!\left(-\alpha^{2}+\varepsilon^{2}-\Gamma^{2}\omega_{0}^{2}\right)
+\Bigl[-\Gamma^{2}(\varepsilon^{2}-\alpha^{2})(T_c-T_h)
+m^{2}\varepsilon((\alpha+\varepsilon)T_c+ (\alpha-\varepsilon)T_h)\omega_{0}^{2}
\\
&
-\Gamma^{2}(\varepsilon^{2}-\alpha^{2})(T_c-T_h)
+ m^{2}\varepsilon\big((\alpha+\varepsilon)T_c+(\alpha-\varepsilon)T_h\big)\omega_{0}^{2}
+ m^{2}\Gamma^{2}(-T_c+T_h)\omega_{0}^{4}\Bigr]\omega_c
- m^{2}\Gamma\big((\alpha+\varepsilon)T_c\\ &
+(\alpha-\varepsilon)T_h\big)\omega_{0}^{2}\omega_c^{2}
\Biggr\} 
-2\alpha\varepsilon_{0}^{2}t_c \Biggl\{
\Gamma(m+\Gamma t_c)\!\left(\alpha^{2}-\varepsilon^{2}+\Gamma^{2}\omega_{0}^{2}\right)\!(m+\Gamma t_c+2m t_c^{2}\omega_{0}^{2})- m\varepsilon\Bigl[m(\alpha^{2}-\varepsilon^{2})t_c^{2}
\\
&+\big(m^{3}+2m^{2}\Gamma t_c+m\Gamma^{2}t_c^{2}-2\Gamma^{3}t_c^{3}\big)\omega_{0}^{2}
+ m^{3}t_c^{2}\omega_{0}^{4}\Bigr]\omega_c+ m^{2}\Bigl[\Gamma(\alpha^{2}-\varepsilon^{2})t_c^{2}
+\big(m^{2}\Gamma+2\Gamma^{3}t_c^{2}+2m t_c(\Gamma^{2}-\varepsilon^{2}t_c^{2})\big)\omega_{0}^{2}
\\
&+2m\Gamma t_c^{2}(m+\Gamma t_c)\omega_{0}^{4}\Bigr]\omega_c^{2}
- m^{3}\varepsilon\,t_c^{2}(m-2\Gamma t_c)\omega_{0}^{2}\omega_c^{3}
+ m^{4}\Gamma\,t_c^{2}\omega_{0}^{2}\omega_c^{4}
\Biggr\}.
\end{split}
\end{equation}

\begin{equation}
\begin{split}
A_x =&
\Gamma \Delta_1  \biggl\{
T_h \big(
-(\alpha^2 - \varepsilon^2)(\Gamma^3 + \varepsilon m^2 \omega_c)
+ \omega_0^4(\Gamma m^4 \omega_c^2 + 2\Gamma^3 m^2)
+ \Gamma  m^2 \omega_0^2(\alpha^2 - \varepsilon^2)
\big)+ T_c \big(
\Gamma m^2 \omega_c^2(2\varepsilon(\alpha+\varepsilon)\\ &+m^2\omega_0^4)
- \varepsilon m^2(\alpha+\varepsilon)^2\omega_c
+ \Gamma (\alpha+\varepsilon)^2(\Gamma^2 + m^2 \omega_0^2)
\big)
\biggl\}+ 2\xi_0^2 t_c \biggl\{
m t_c^3 \big(
(\alpha+\varepsilon)^2 + m^2\omega_0^4
\big)
\big(
\Gamma^2(\alpha^2-\varepsilon^2)\\ &
+ \varepsilon\omega_c(\Gamma^3 + m^2 \omega_c(\Gamma\omega_c-\varepsilon))
+ \omega_0^2(\Gamma^4 + \Gamma^2 m^2\omega_c^2)
\big)+ t_c^2 \big(
m^2 \big[
\omega_0^2(\alpha+\varepsilon)(\Gamma^3(2\alpha+\varepsilon)
+ m^2\omega_c(\Gamma(2\alpha+\varepsilon)\omega_c - 2\alpha\varepsilon))
\\ &
+ \varepsilon(\alpha+\varepsilon)\omega_c(
\omega_c(2\Gamma^3 + m^2\omega_c(\Gamma\omega_c - \varepsilon))
- \Gamma^2\varepsilon)+ \omega_0^4 \big(
\Gamma^5
+ m^2\omega_c(\omega_c(2\Gamma^3 + m^2\omega_c(\Gamma\omega_c-\varepsilon))
- \Gamma^2\varepsilon)
\\ &+ \alpha\Gamma m^2(\alpha+\varepsilon)
\big)+ \Gamma m^2 \omega_0^6(\Gamma^2 + m^2\omega_c^2)
\big]+ \Gamma(\alpha+\varepsilon)(\Gamma^4\varepsilon + \alpha m^2(\alpha^2 - \varepsilon^2))
\big)+ 2\Gamma m t_c \big(
\Gamma^3\varepsilon(\alpha+\varepsilon)
+ m^2\big(
\alpha\Gamma\omega_0^2(\alpha+\varepsilon)
\\ &+ \varepsilon(\alpha+\varepsilon)\omega_c(\Gamma\omega_c - \alpha)
+ \omega_0^4(\Gamma^3 + \Gamma m^2\omega_c^2)
\big)
\big)
+ m^4\big(
\alpha\Gamma\omega_0^2(\alpha+\varepsilon)
+ \varepsilon(\alpha+\varepsilon)\omega_c(\Gamma\omega_c - \alpha)
+ \omega_0^4(\Gamma^3 + \Gamma m^2\omega_c^2)
\big)
\\ &+ \Gamma^3 \varepsilon m^2 (\alpha+\varepsilon)
\biggl\}.
\end{split}
\end{equation}

\begin{equation}
\begin{split}
A_y &=
\Gamma \Delta_1   \Bigl[
\Gamma m^2 \omega_0^2 \omega_c^2 (T_c+T_h)
+\varepsilon m^2 \omega_c^3 (T_c+T_h)
+\Gamma T_c \left(\alpha^2 + 2\Gamma^2 \omega_0^2 - \varepsilon^2\right)
+2\Gamma^2 \varepsilon T_c \omega_c
+\Gamma (\varepsilon - \alpha)^2 T_h
\Bigr]
\\
&
+2\xi_0^2 t_c \Bigl[
\varepsilon m^3 \omega_c^3 (\Gamma t_c + m)
+ m^2 \omega_c^2 \bigl(
\Gamma m^2 \omega_0^4 t_c^2
+ \varepsilon t_c \left(\Gamma (\varepsilon - \alpha)t_c + \varepsilon m\right)
+ \Gamma m \omega_0^2(\Gamma t_c + m)
\bigr)+ \Gamma^3 m^2 \omega_0^4 t_c^2
+ \Gamma m \omega_0^2 \bigl[
m(\Gamma^2 \\
&+ \alpha(\alpha - \varepsilon)t_c^2)
+ \Gamma^3 t_c
\bigr]+ \varepsilon m \omega_c \bigl[
m(\Gamma^2 + \alpha(\varepsilon - \alpha)t_c^2)
+ \Gamma^3 t_c
\bigr]
+ \Gamma(\varepsilon - \alpha)(\Gamma t_c + m)(\Gamma \varepsilon t_c - \alpha m)
\Bigr].
\end{split}
\end{equation}
\end{widetext}

\end{document}